\documentclass[final,1p,times]{elsarticle} 
\usepackage{graphicx} 
\usepackage{subfigure}
\usepackage{amssymb} 
\usepackage{amsthm} 

\journal{Nuclear Physics A} 
\begin{document} 

\begin{frontmatter} 


\title{Lessons from PHOBOS}

\author{Wit Busza (for the PHOBOS \footnote{For the full list of the
PHOBOS collaboration and acknowledgments, see appendix ``Collaborations'' of this volume.} collaboration)}

\address{Massachusetts Institute of Technology,
77 Massachusetts Ave, Cambridge MA 02139}

\begin{abstract} 
In June 2005 the PHOBOS Collaboration completed data taking at RHIC.  In five years of operation PHOBOS	recorded information for Au+Au at $\sqrt{s_{NN}}$ = 19.6, 62.4, 130, and 200 GeV, Cu+Cu at 22.4, 62.4 and 200 GeV, d+Au at 201 GeV, and p+p at 200 and 410 GeV, altogether more than one billion collisions.  Using these data we have studied the energy and centrality dependence of the global properties of charged particle production over essentially the full 4$\pi$ solid angle and (for pions near mid rapidity) charged particle spectra down to transverse momenta below 30 MeV/c.  We have also studied correlations of particles separated in pseudorapidity by up to 6 units.  We find that the global properties of heavy ion collisions can be described in terms of a small number of simple dependencies on energy and centrality, and that there are strong correlations between the produced particles.  To date no single model has been proposed which describes this rich phenomenology.  In this talk I summarize what the data is explicitly telling us.
\end{abstract} 

\end{frontmatter} 



\vspace{-10mm}

\begin{figure}[htbp]
\centering
\subfigure[]{
    \includegraphics[width=.48\linewidth]{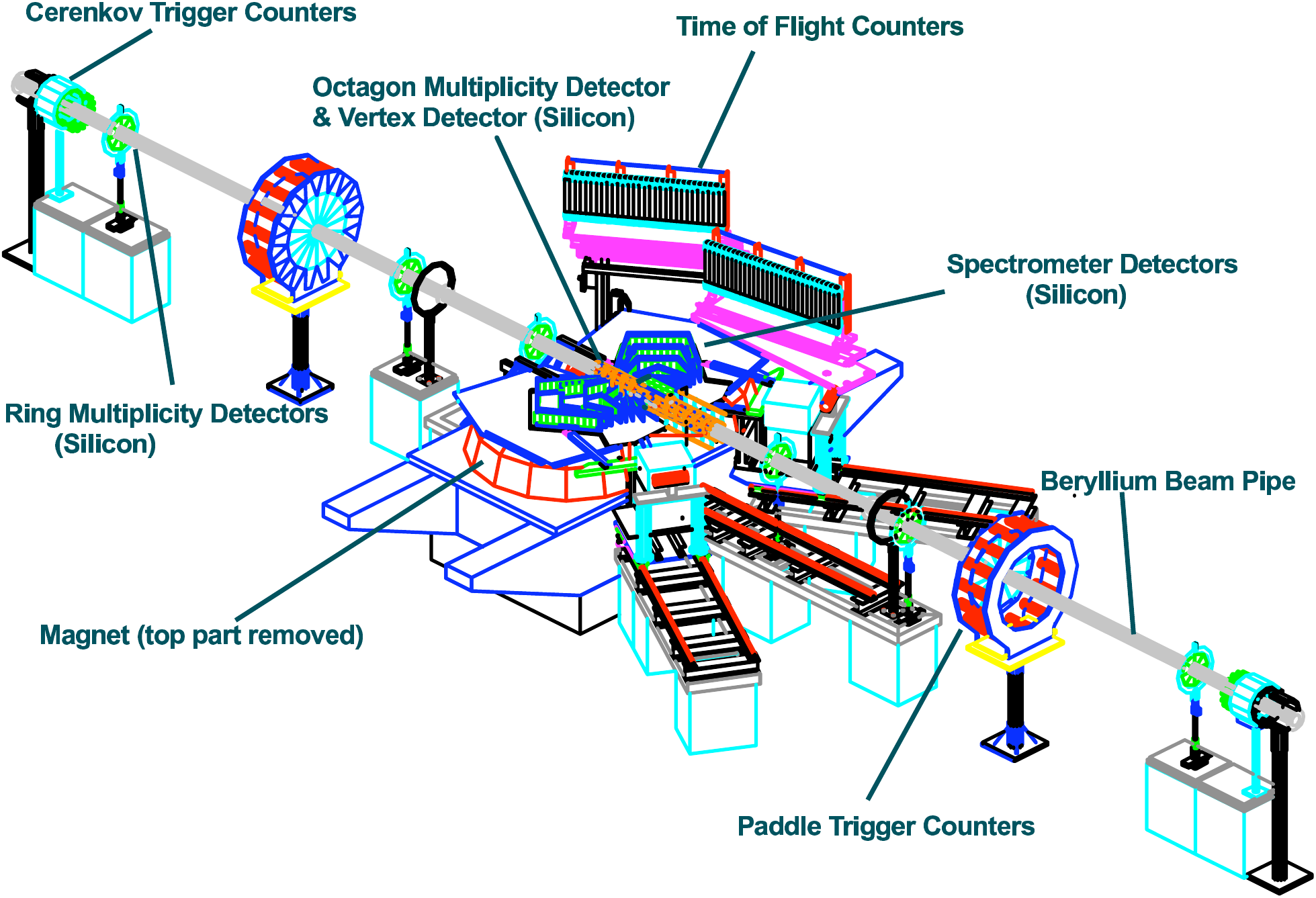}
    \label{figure 1a}
}
\hfill
\subfigure[]{
    \includegraphics[width=.48\linewidth]{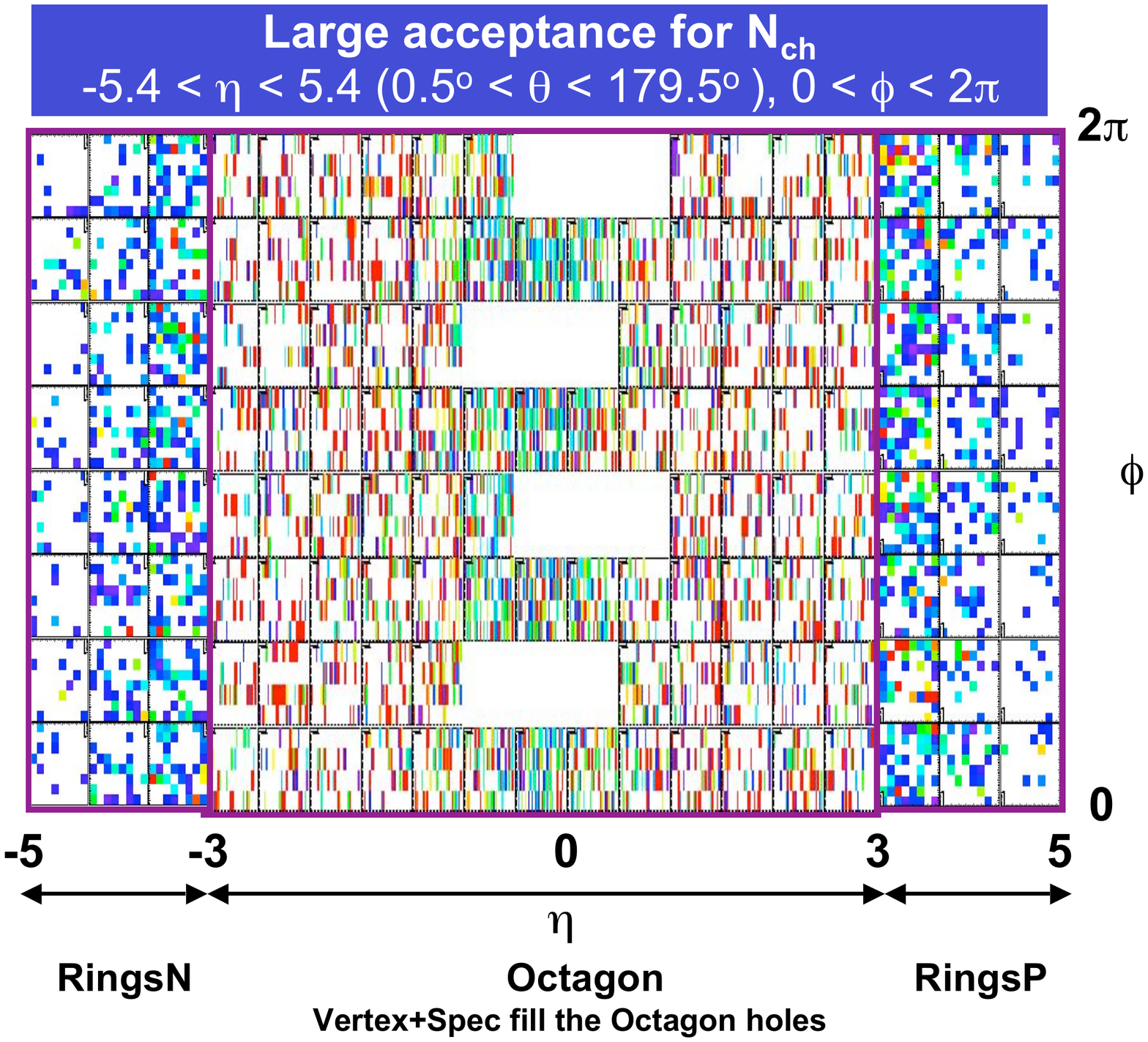}
    \label{figure 1b}
}
\vspace{-3mm}
\caption{(Color online) The PHOBOS detector and acceptance for charged particles.}
\end{figure}

In 2005 PHOBOS stopped taking data.  Although we are still analysing and publishing results obtained in the first five RHIC runs it is fair to state that the PHOBOS project is winding down and that it is a perfect time to review and assess what PHOBOS has taught us.  Given the space constraint, rather than listing and discussing all our results and achievements in encyclopedic fashion, I will focus on what it is that we have learnt to date from the PHOBOS data, in particular from those data which we were able to  obtain as a consequence of the strengths of PHOBOS - the broadest pseudorapidity coverage and lowest transverse measurement capability of all RHIC experiments.  The results of our studies published to date can be found in ref. 1-37.

My talk is in the form of ``eight lessons learnt from PHOBOS data".  But first, as a reminder, I say a few words about the PHOBOS detector at RHIC.  In essence PHOBOS is an almost 4$\pi$ acceptance, charged particle multiplicity and pseudorapidity ($\eta$) detector, including two small acceptance spectrometers with which charged particle spectra can be measured just forward of mid-rapidity ($0\le\eta\le1.5$ and $\Delta\phi< 0.2$ rad).  A sketch of the PHOBOS detector and its acceptance are given in Fig. 1, and a detailed description of the experiment is available in ref. 38.  As can be seen, the $\eta$-coverage is $\pm$5.4 units.  In the spectrometer, for pions the transverse momentum $(p_T)$ measurement capability extends down to 30 MeV/c and charged hadrons can be identified up to about 3 GeV/c. 

{\underline{{\bf Lesson 1:}} {\bf At RHIC there is no anomalous production of low $p_T$ particles.}

One of the design features of PHOBOS was sensitivity to low transverse momentum ($p_T$) particles.  The rationale for this is straightforward.  The wavelength of produced particles tends to reflect the geometrical dimensions of the source.  An anomalous production of particles with $p_T \le \frac{\hbar}{R}$
($\hbar$ = Planck's constant and R = the radius of the colliding nuclei) would be an indication of the occurrence in heavy ion collisions of phenomena coherent over distances that are large compared to nucleon dimensions (i.e. new phenomena related to the large interacting volumes in heavy ion collisions) and not resulting from the superposition of particles produced in nucleon-nucleon collisions.  A good example of such new phenomena would be the expected large flux of very low $p_T$ pions from the decay of a disoriented chiral condensate (DCC).

\vspace{-2mm}

\begin{figure}[h]
\centering
\subfigure[]{
\includegraphics[width=.45\linewidth]{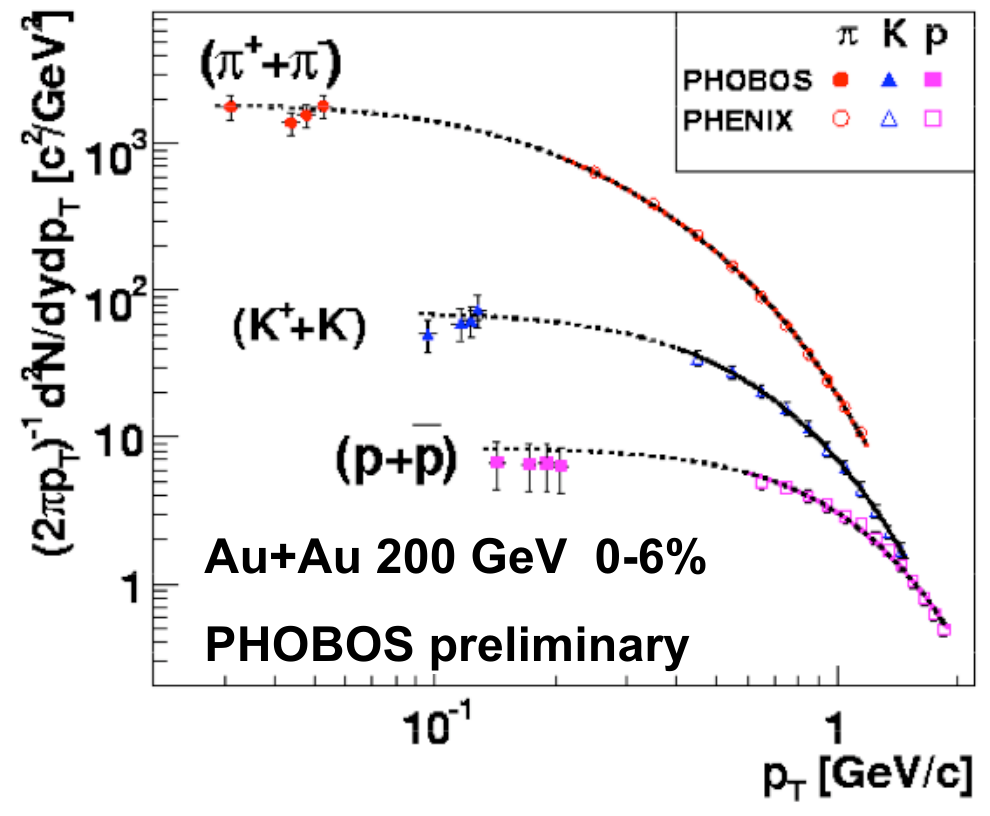}
\label{}
}
\hfill
\subfigure[]{
\includegraphics[width=.45\linewidth]{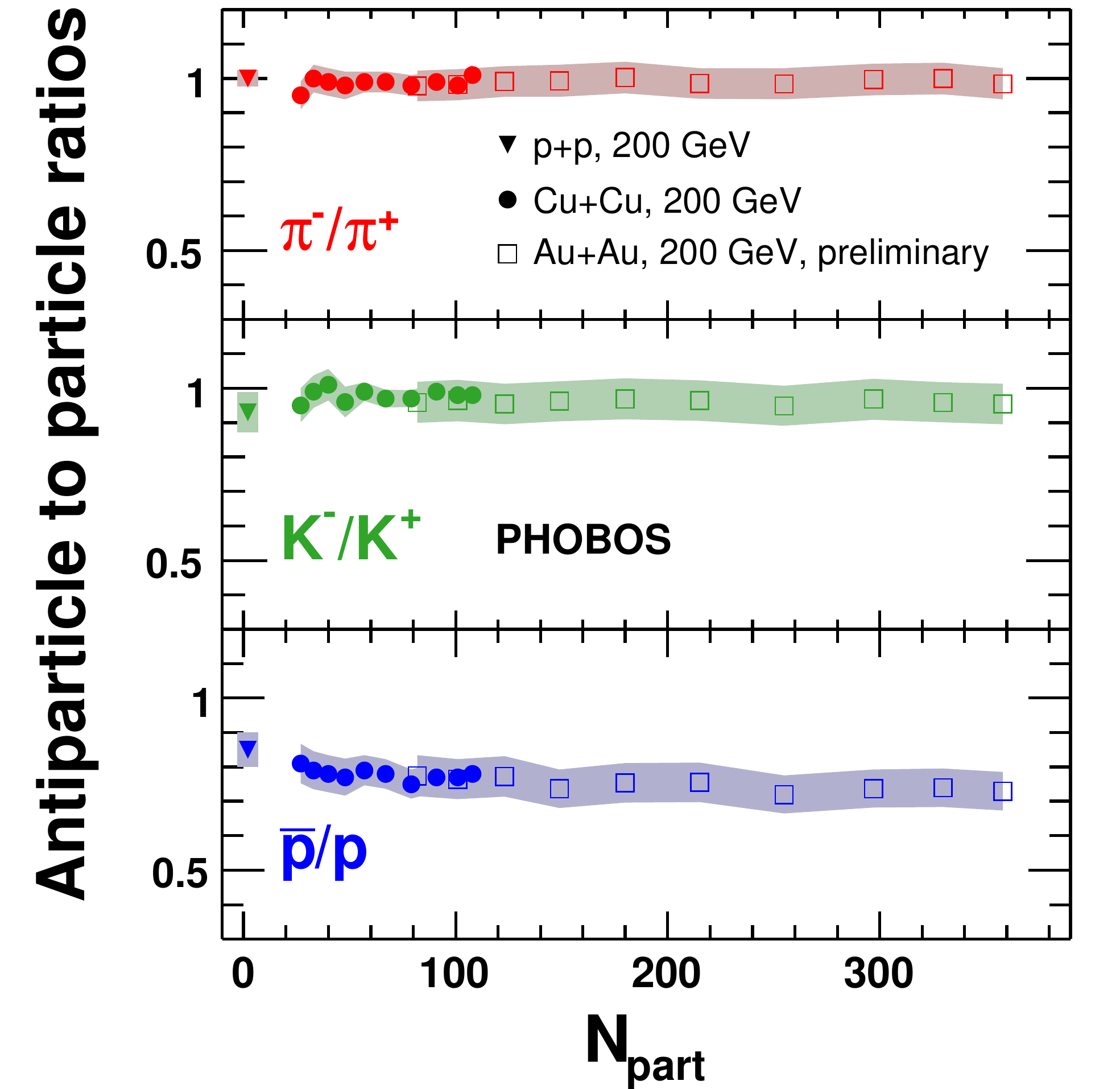}
\label{}
}
\vspace{-2mm}
\caption[]{\label{}
(Color online) (a) Example of particle spectra at low $p_T$ for central Au+Au collisions near midrapidity $(0.2\le\eta\le1.4)$.  The data show no indication of anomalous production of low $p_T$ particles.
(b) Examples of the centrality independence of antiparticle to particle ratios.}
\end{figure}
PHOBOS measured pions, kaons, and protons plus antiprotons down to $p_T \sim$ 30 MeV/c, 100 MeV/c and 150 MeV/c respectively, for Au+Au collisions at $\sqrt{s_{NN}}$=62.4 and 200 GeV~\cite{9,24}. The results are consistent with expectations from a simple extrapolation of particle production fitted at higher transverse momenta using a blast-wave parameterization.  Fig. 2 (a) is an illustration of PHOBOS low $p_T$ data for the most central Au+Au collisions at $\sqrt{s_{NN}}$= 200 GeV.

{\underline{{\bf Lesson 2:}} {\bf Although at RHIC at mid-rapidity a zero net baryon density is not reached, the ratio of particles to antiparticles is already independent of the colliding system size.}

In very general terms, in heavy ion collisions, two sources contribute to the production of baryons at mid-rapidity. One is pair production and the other is ``stopping" of the incident baryon number as the two nuclei interpenetrate and interact with each other.  Only the first of these sources contributes to the production of antibaryons.  Naively one would thus expect that these two very different production mechanisms would depend differently on the thickness of the colliding nuclear material, i.e. on the impact parameter or number of participants ($N_{part}$), leading to an antiparticle/particle ratio that is centrality dependent.  This is not what is observed by PHOBOS~\cite{4, 26}.  For example, Fig. 2 (b) shows a nearly constant $\frac{\pi^-}{\pi^+}$,  $\frac{k^-}{k^+}$ and $\frac{\bar{p}}{p}$ ratio as a function of $N_{part}$ for Cu+Cu and Au+Au collisions at $\sqrt{s_{NN}}$=200 GeV.  An interesting related fact is that the net proton yield at midrapidity is proportional to $N_{part}$~\cite{9}.

{\underline{{\bf Lesson 3:}} {\bf The production process shows signs of significant saturation during the early stages of the collision.}

From the beginning of the RHIC experimental program PHOBOS has carried out a very detailed and systematic study of pseudorapidity and azimuthal distributions for all available colliding systems, at all available energies, and for a large variety of impact parameters (of $N_{part}$)~\cite{5, 8, 15, 16}.  The resultant extensive set of data allows investigation of the trends and systematics of the multiparticle production process.

\begin{figure}[htbp] 
   \centering
   \includegraphics[width=.4\linewidth]{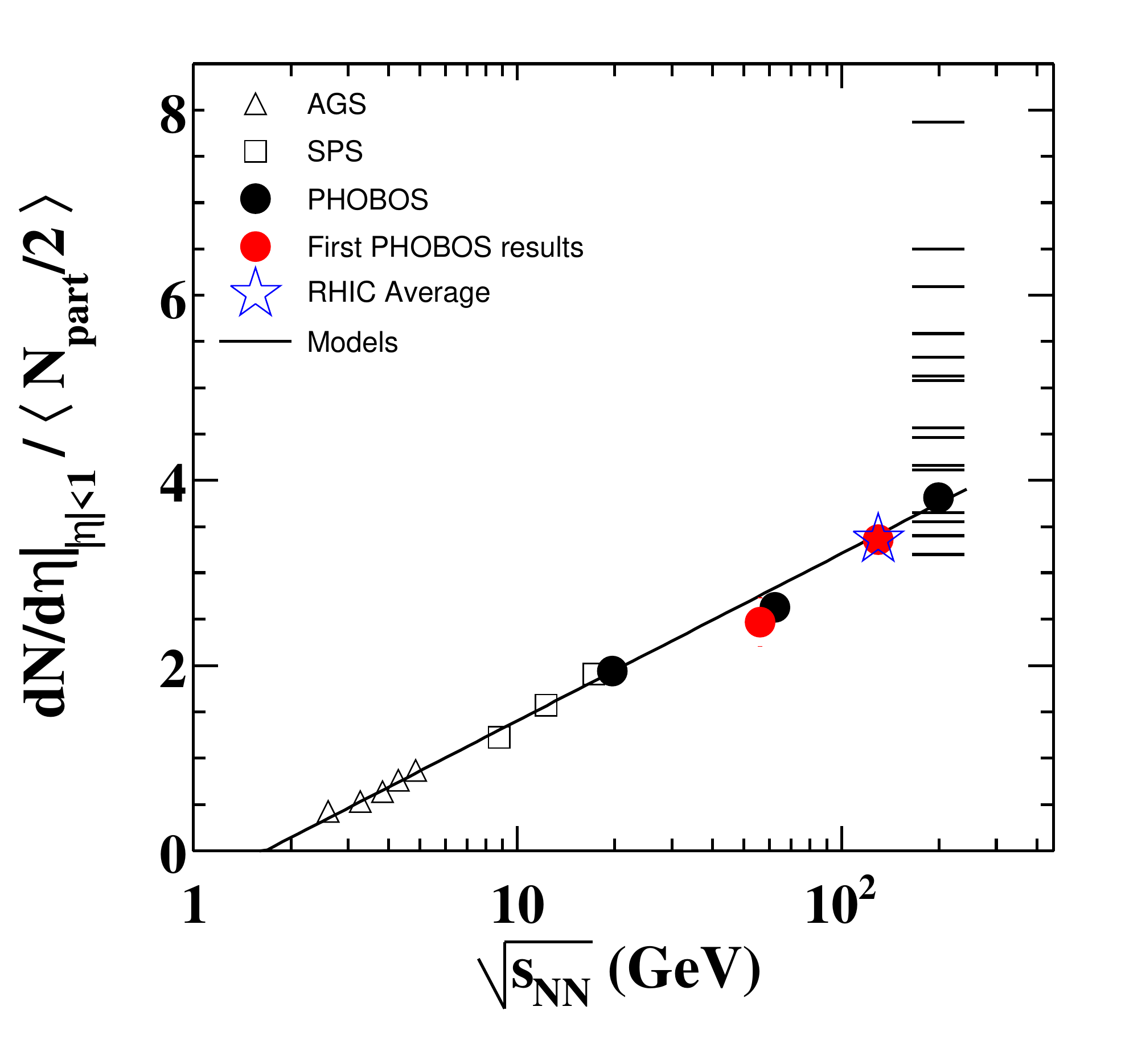} 
   \vspace{-3mm}
   \caption{(Color online) The first results at RHIC (from PHOBOS) on midrapidity particle density compared to predictions of models (indicated schematically by horizontal lines).}
   \label{figure 4}
\end{figure}

\begin{figure}[h!]
\centering
\subfigure[]{
\includegraphics[width=.50\linewidth]{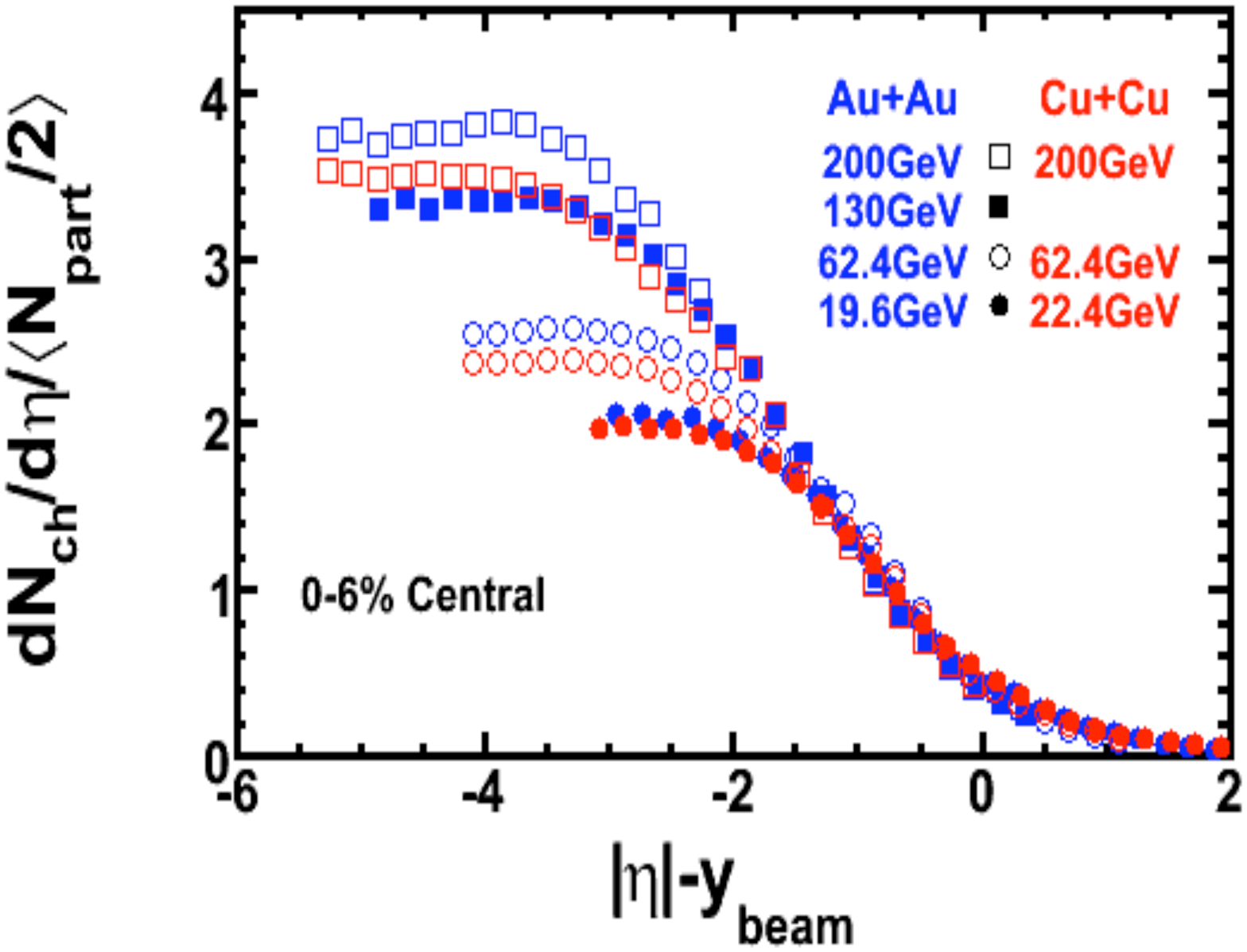}
\label{}
}
\hfill
\subfigure[]{
\includegraphics[width=.45\linewidth]{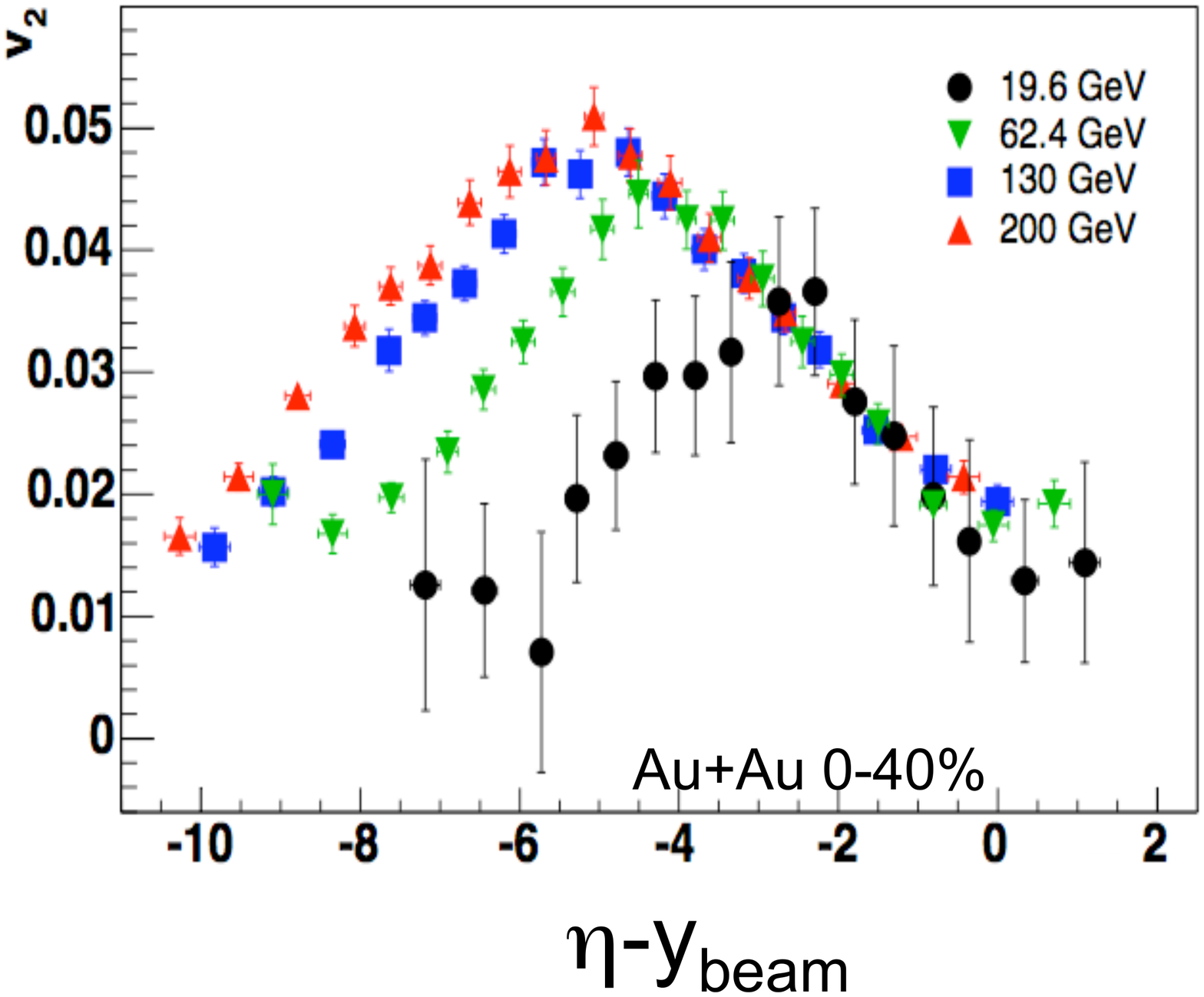}
\label{}
}
\vspace{-2mm}
\caption[]{\label{figure7}
(Color online) (a) $\frac{dN}{d\eta}$ and (b) $v_2$ plotted in the rest frame of one of the incident nuclei.  These are examples of ``limiting fragmentation" and ``extended longitudinal scaling".  See refs. 5, 8, 14, 15, 16.}
\end{figure}
\begin{figure}[h!]
\centering
\subfigure[]{
\includegraphics[width=.45\linewidth]{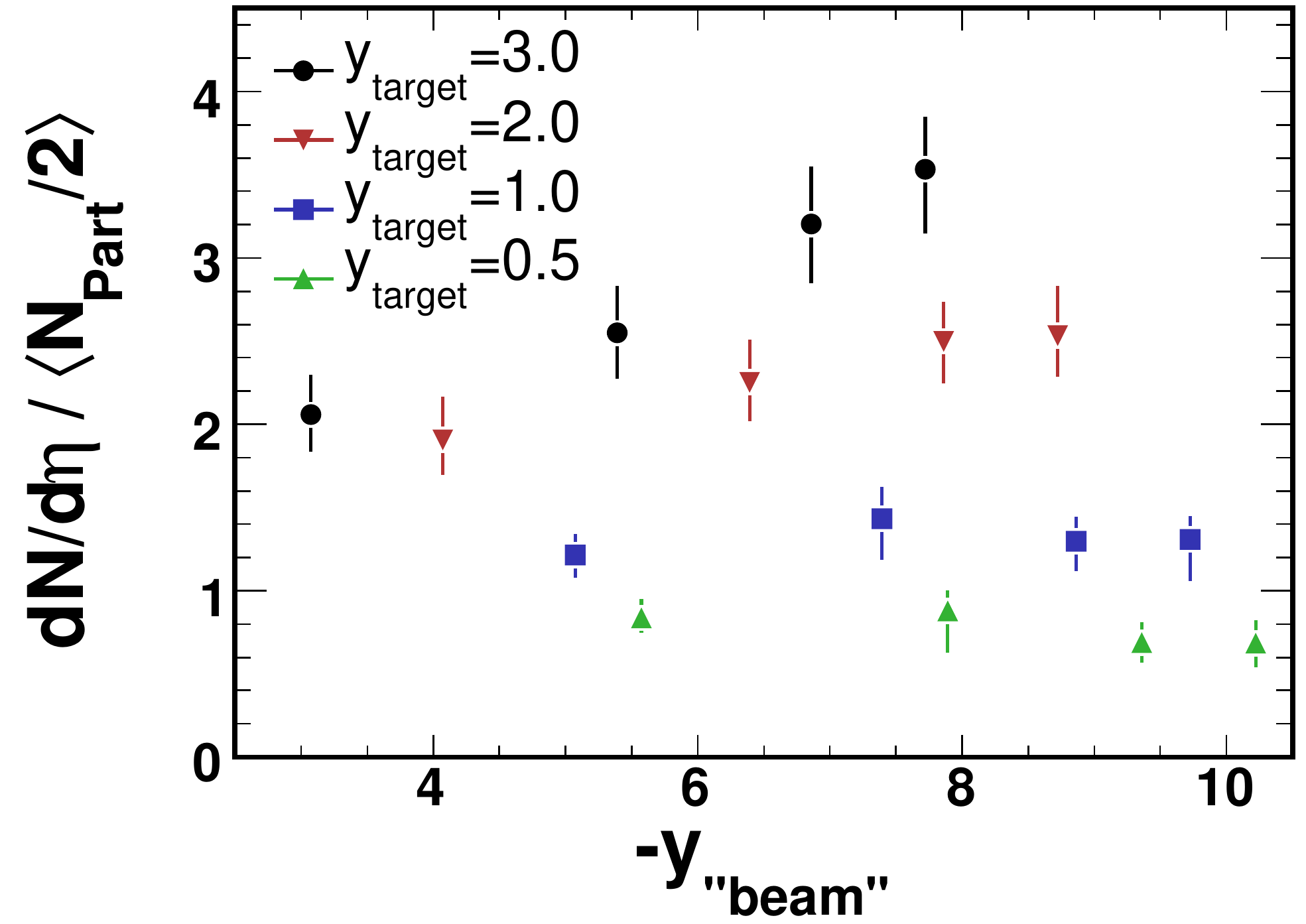}
\label{}
}
\hfill
\subfigure[]{
\includegraphics[width=.45\linewidth]{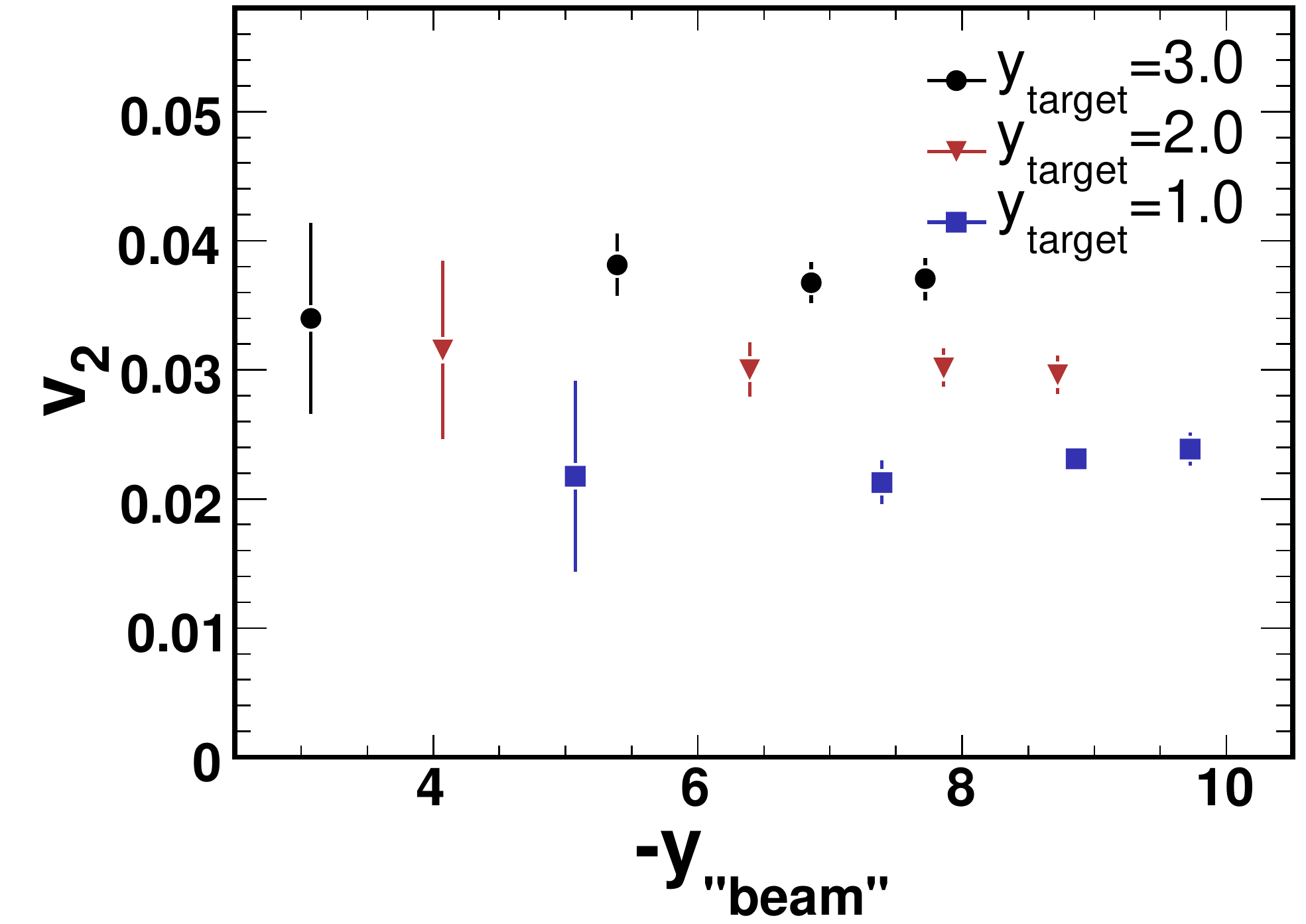}
\label{}
}
\vspace{-2mm}
\caption[]{\label{}
(Color online) (a) $\frac{dN}{d\eta}$ and (b) $v_2$  as seen at $y=0$, plotted in a frame where one incident nucleus has rapidity $y_{target}$ and the other $y_{``beam"}$.  These are direct  evidence of some kind of saturation in the production process.  The data is the same as in Fig. 4.  Note: a set of points with a given $y_{target}$ in Fig. 5 corresponds to a set of points with a given $|\eta|- y_{beam}$ in Fig. 4.
}
\end{figure}

The very first physics result at RHIC (submitted for publication by PHOBOS 5 weeks after the first collision at RHIC~\cite{37}) turned out to be a surprise to most theorists.  The particle density at midrapidity was significantly lower than most expectations.  Fig. 3 shows a compilation of the observed energy dependence of the midrapidity particle density.  It also shows schematically the predictions of the many theoretical models which predated RHIC results~\cite{15}.  We note that it is the models which include saturation that are in best agreement with the data, suggesting that some kind of saturation plays an important role in heavy ion collisions.

Direct evidence that saturation occurs in heavy ion collisions can be seen in a careful study of the PHOBOS data on the energy dependence of the pseudorapidity distributions.

Assuming that the pseudorapidity $\eta$ is a good approximation for the rapidity $y$ we can use the extensive PHOBOS data on $\frac{dN}{d\eta}$ and elliptic flow $v_2$ to reconstruct the density of particles $\frac{dN}{d\eta}|_{y=0}$ and the elliptic flow $v_2|_{y=0}$ in asymmetric (in energy) collisions of nuclei.  For example at y=0,  in a frame where one nucleus (the ``target" nucleus) is at rest, we find that the particle density and $v_2$ are independent of the rapidity of the other nucleus (the ``beam" nucleus).  This is the well known phenomenon of ``limiting fragmentation", see Fig. 4.  This saturation of particle production and elliptic flow occurs not only for ``target" nuclei at rest.  For any rapidity $y_{target}$ of a ``target" nucleus, we find that, provided that $y_{``beam"}$ is above some $y_{target}$-dependent threshold, $\frac{dN}{d\eta}|_{y=0}$ and $v_2|_{y=0}$ are independent of $y_{``beam"}$. This for example is evident in Fig. 5.  Furthermore, the threshold increases with $y_{target}$.  In other words, if an asymmetric beam facility was constructed and one beam had a fixed rapidity $y_{target}$, $\frac{dN}{d\eta}|_{y=0}$ and $v_2|_{y=0}$ would increase as one increased $y_{beam}$ until a saturation value was reached.  From then on no further increase would be seen with increase of ``$y_{beam}$".  The only way to further increase $\frac{dN}{d\eta}|_{y=0}$ and $v_2|_{y=0}$ would be to increase $y_{target}$.

This phenomenon, named by PHOBOS ``extended longitudinal scaling", is a direct manifestation that some kind of saturation of the particle multiplicity, of the directed flow (not discussed here)~\cite{13} and of the elliptic flow occur in heavy ion collisions.

{\underline{{\bf Lesson 4:}} {\bf In heavy ion collisions, energy and system dependence factorize.}}

In nucleon-nucleon collisions, the fraction of the cross-section which gives rise to hard scattering increases with energy.  Since in heavy ion collisions the number of soft collisions is proportional to $N_{part}$ and the number of hard collisions to $N_{coll}$, we would not expect the energy and system dependences to factorize.  Surprisingly, the global features studied by PHOBOS do factorize.  The energy dependence of the global features of the data is independent of the colliding system size, and the system size dependence  or $N_{part}$ dependence of the global features is independent of the energy.  This factorization is seen for example in the total production of charged particles~\cite{5, 14, 15, 16, 40} (see Fig. 6), for the midrapidity particle density~\cite{3}, for the particle production in the fragmentation regions~\cite{5, 40}, and even for the production of particles of definite transverse momentum~\cite{3}.

\begin{figure}[h]
\centering
\subfigure[]{
\includegraphics[width=.45\linewidth]{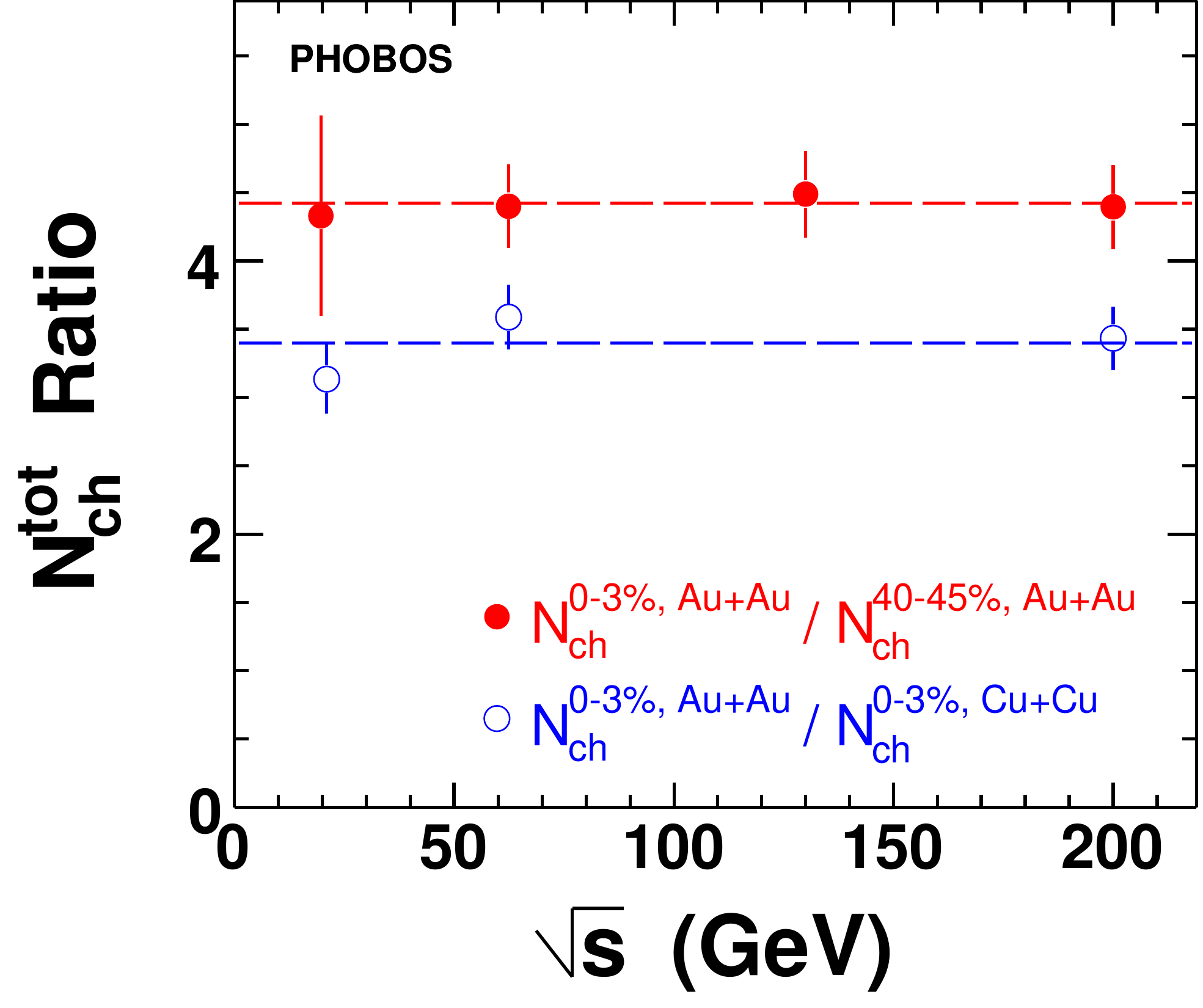}
\label{}
}
\hfill
\subfigure[]{
\includegraphics[width=.45\linewidth]{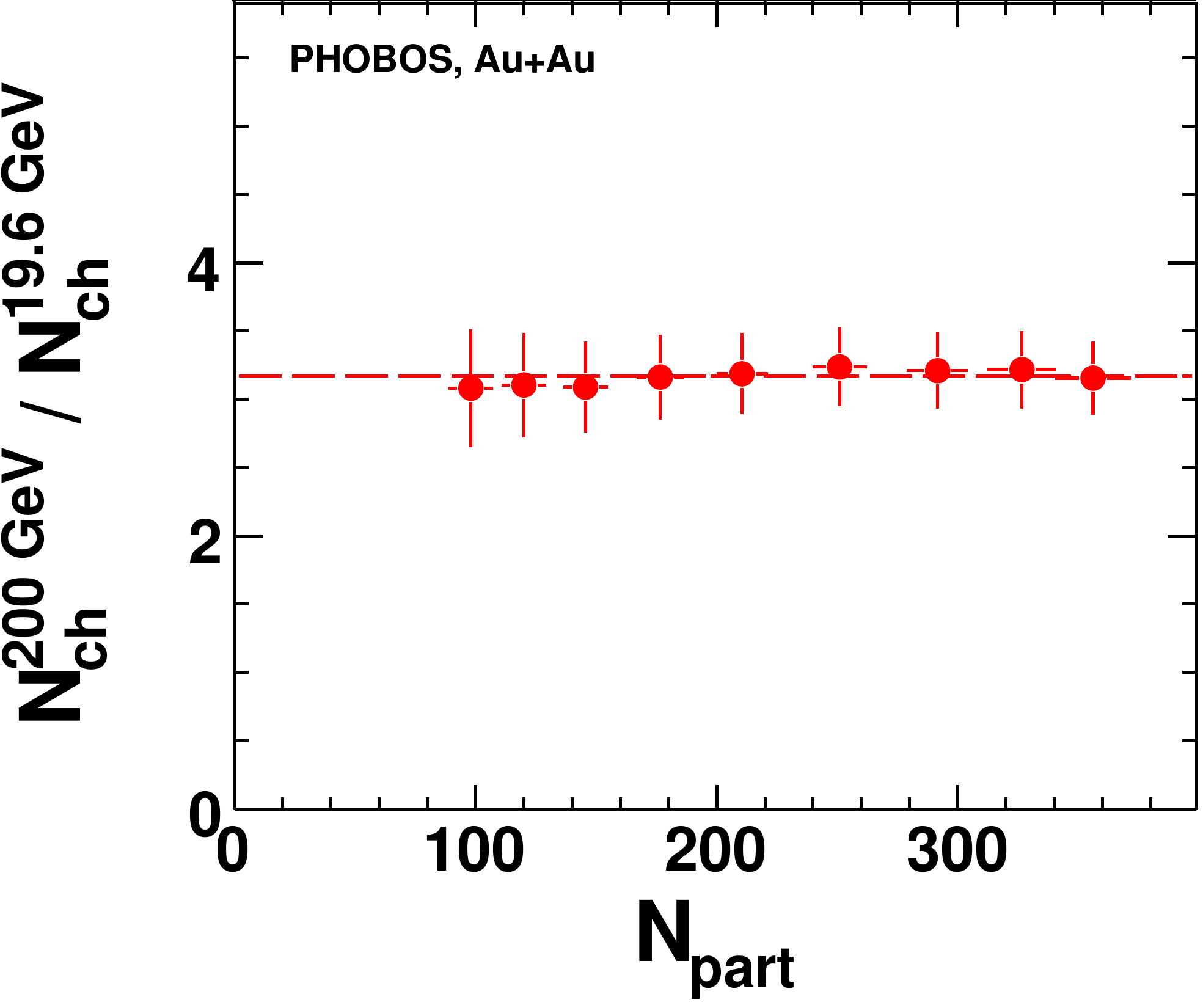}
\label{}
}
\vspace{-2mm}
\caption[]{\label{}
(Color online) Ratios of the total charged particles produced in collisions of various systems.  These are good examples of the energy and system dependence factorization.}
\end{figure}

{\underline{{\bf Lesson 5:}} {\bf The total particle production in $e^++e^-$, p+p, p+A, and A+A collisions is insensitive to the colliding systems.}
\begin{figure}[h]
\centering
\subfigure[]{
\includegraphics[width=.48\linewidth]{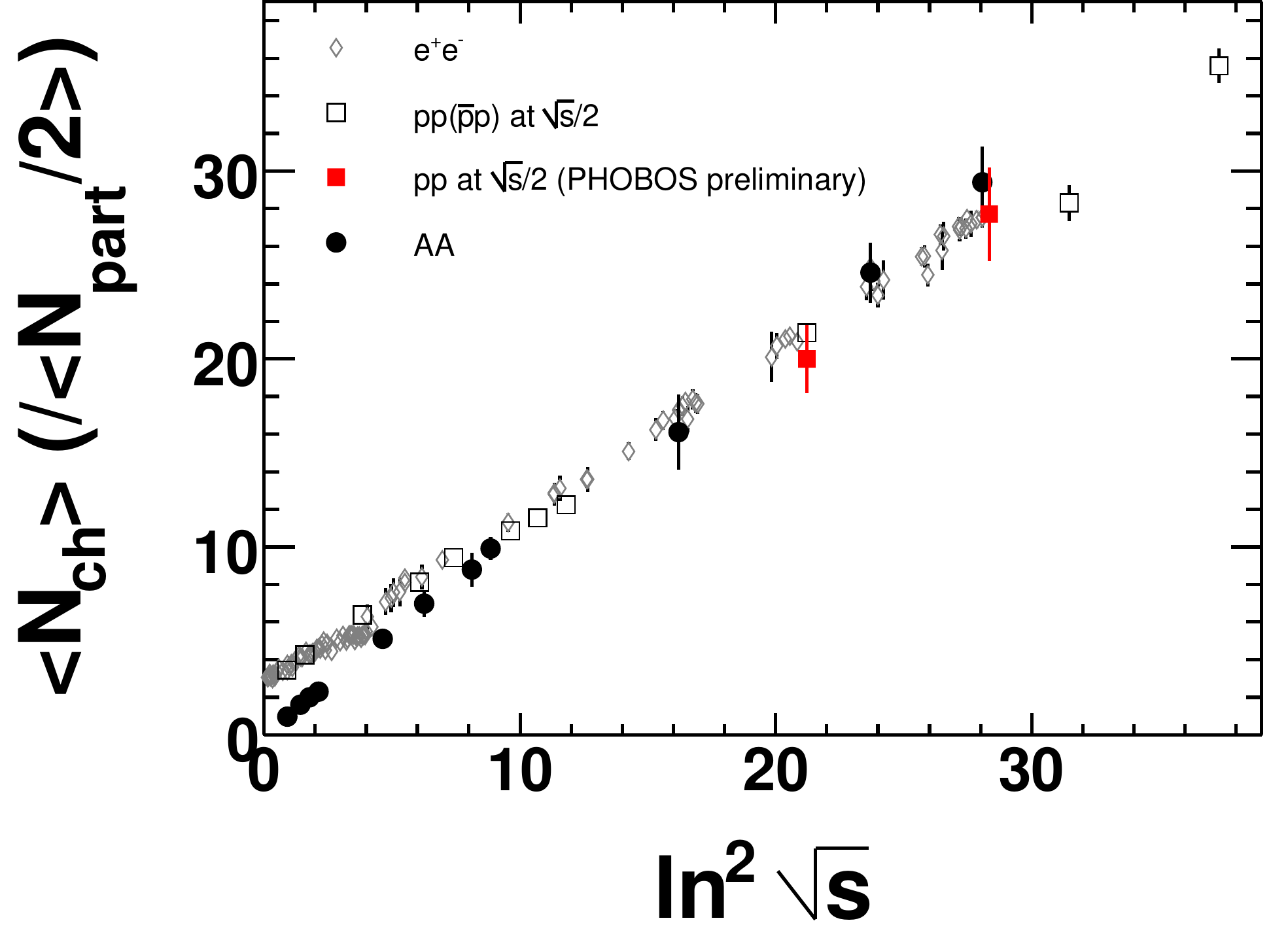}
\label{}
}
\hfill
\subfigure[]{
\includegraphics[width=.43\linewidth]{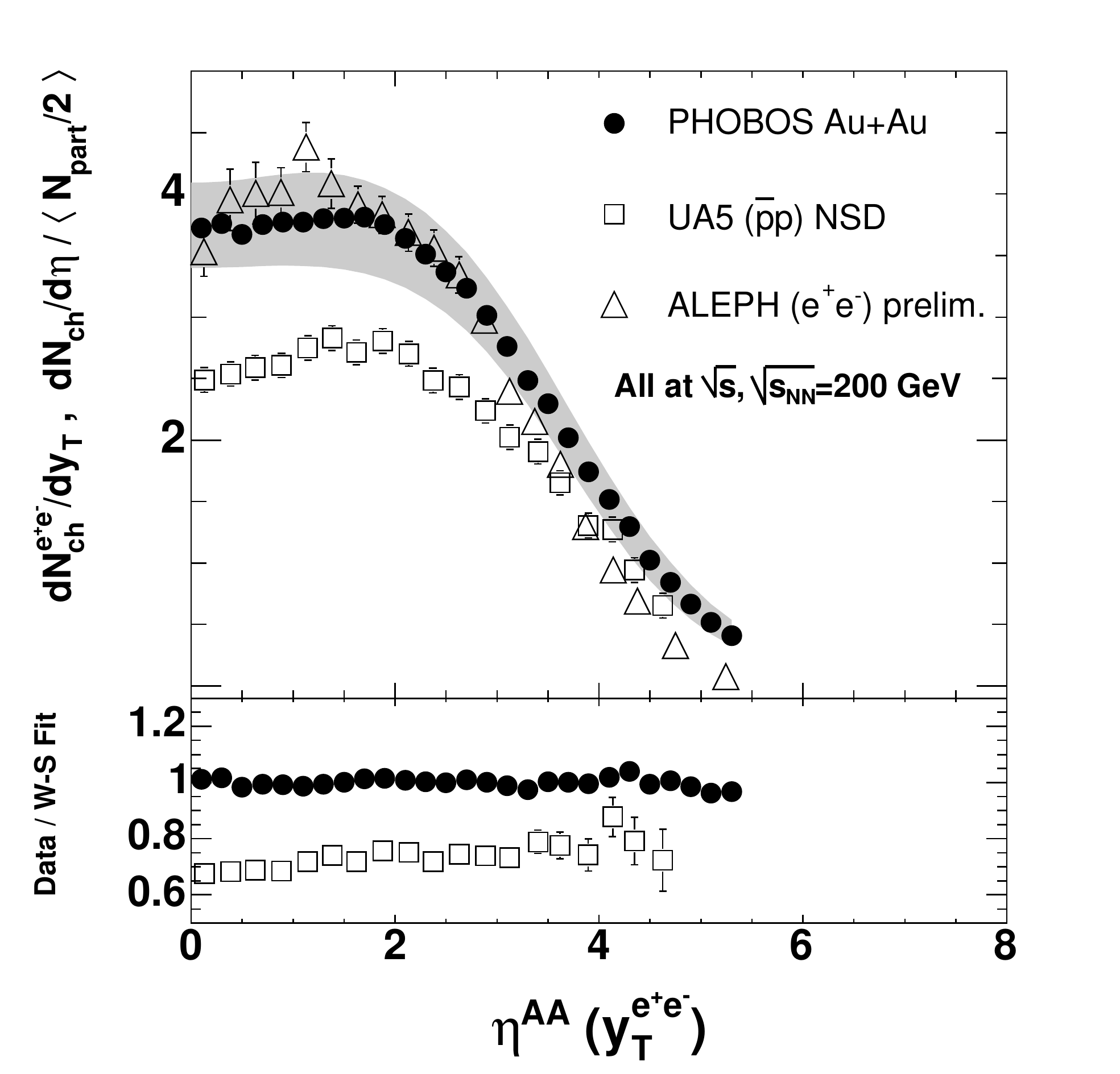}
\label{}
}
\vspace{-2mm}
\caption[]{\label{}
(Color online) Comparison of particle production in $e^++e^-$, p+p, and A+A collisions (a) energy dependence over the range from 2-900 GeV (b) rapidity distributions at 200 GeV.}
\end{figure}

In our current understanding of A+A collisons the intermediate state is very different for high and low energy collisions.  Similarly for our understanding of $e^++e^-$, p+p, p+A and A+A collisions.  Nevertheless the observed trends in the global features of multiparticle production in the collision of these various systems is remarkably similar.  See for example (Fig. 7) the energy and centrality dependence of the total multiplicity and a comparison of the pseudorapidity distributions~\cite{10, 40}.  It is an intriguing question what is it in the mechanism of the collision process that makes the resultant particle production so insensitive to the intermediate state.

{\underline{{\bf Lesson 6:}} {\bf There is clear evidence of a non-trivial correlation between particles separated by large rapidities.}

Hadrons with very different rapidities are produced at large separations in space-time.  Any correlation between such particles, by necessity, must have its origin at early time.  Thus, long range rapidity correlations give valuable information about the very early stages of the collision process.  For this reason PHOBOS has extensively studied triggered two-particle correlations over the uniquely broad longitudinal acceptance of the PHOBOS detector~\cite{1}. To be more specific, we have studied correlations between a charged particle produced near mid-rapidity with $p_{T}>$2.5 GeV/c and a second charged particle with $p_T> 7-35$ MeV/c and separated from the first by up to 4 units in pseudorapidity and $\pm\pi$ in azimuthal angle.  A broadening of the away-side azimuthal correlation compared to elementary collisions is observed at all $\Delta\eta$.  As in p+p collisions, the near-side is characterized by a peak of correlated partners at small angle relative to the trigger.  In central Au+Au collisions an  additional interesting correlation, known as the ``ridge", is found to extend to at least $|\Delta\eta|\sim 4$.  These results are presented and discussed by G. Stephans in a companion talk at this conference~\cite{41}.

{\underline{{\bf Lesson 7:}} {\bf In RHIC collisions, the final state decays into clusters which break up into a large centrality dependent number of particles, covering a broad range of rapidities and azimuthal angles.}  

In PHOBOS, in addition to triggered two-particle correlations, we have studied inclusive two-particle angular correlations~\cite{2, 6, 11}.  We find that some features of the correlation functions in heavy ion collisions are similar to those found in p+p, allowing a similar interpretation in terms of clusters.  In heavy ion collisions, we find a non-trivial decrease in effective cluster size with increasing centrality.  Extrapolating the measured cluster parameters to the full phase space using an independent cluster model we find the surprising result that the effective cluster size and width increase in magnitude to a level ($\ge$ 8 and therefore several GeV effective mass) which seems to challenge most conventional scenarios of the hadronization process.  For details and further discussion of these results see ref. 41.

{\underline{{\bf Lesson 8:}} \bf The final distributions of particles often reflect the ``geometry" of the colliding systems at the instant of collision, rather than the size of the system.}

Naively, one would expect that in heavy ion collisions two parameters would be of paramount importance - total energy and some measure of how much overlapping matter is involved in the two colliding systems.  It comes as a surprise to us that for many observables it is not $N_{part}$ or $N_{coll}$ which determine the outcome of the collision but rather the geometrical shape of the overlapping region at the instant of collision.  By comparing Cu+Cu with Au+Au collisions we find, for example, that the cluster size (see lesson 7 above) or the particle production in the fragmentation region depends on the shape rather than volume of the overlapping nuclei~\cite{2, 5}.  Another, observation made by PHOBOS, is that the elliptic flow parameter $v_2$ observed for Cu+Cu and Au+Au collisions can be meaningfully compared only if one takes into account the event-by-event geometrical or eccentricity fluctuations (participant eccentricity)~\cite{7, 8, 39}.

To conclude, in addition to significant contributions to the discoveries and evolution of the current picture of RHIC physics~\cite{15}, PHOBOS data has revealed some intriguing features.  The significance of most of these features is still not well understood.  It will be interesting and instructive to see if the features seen in PHOBOS data at RHIC will continue to be prominent at LHC.  To facilitate a quick comparison, as a conclusion to my talk, I have taken some of the PHOBOS data and made a linear extrapolation to the LHC energies~\cite{42}. The extrapolated results for LHC, $\frac{dN(ch)}{d\eta} \sim 1100$ and $v_{2} \sim 0.075$ at midrapidity, are shown below in Fig. 8.

\vspace{-4mm}

\begin{figure}[h]
\centering
\subfigure[]{
\includegraphics[width=.52\linewidth]{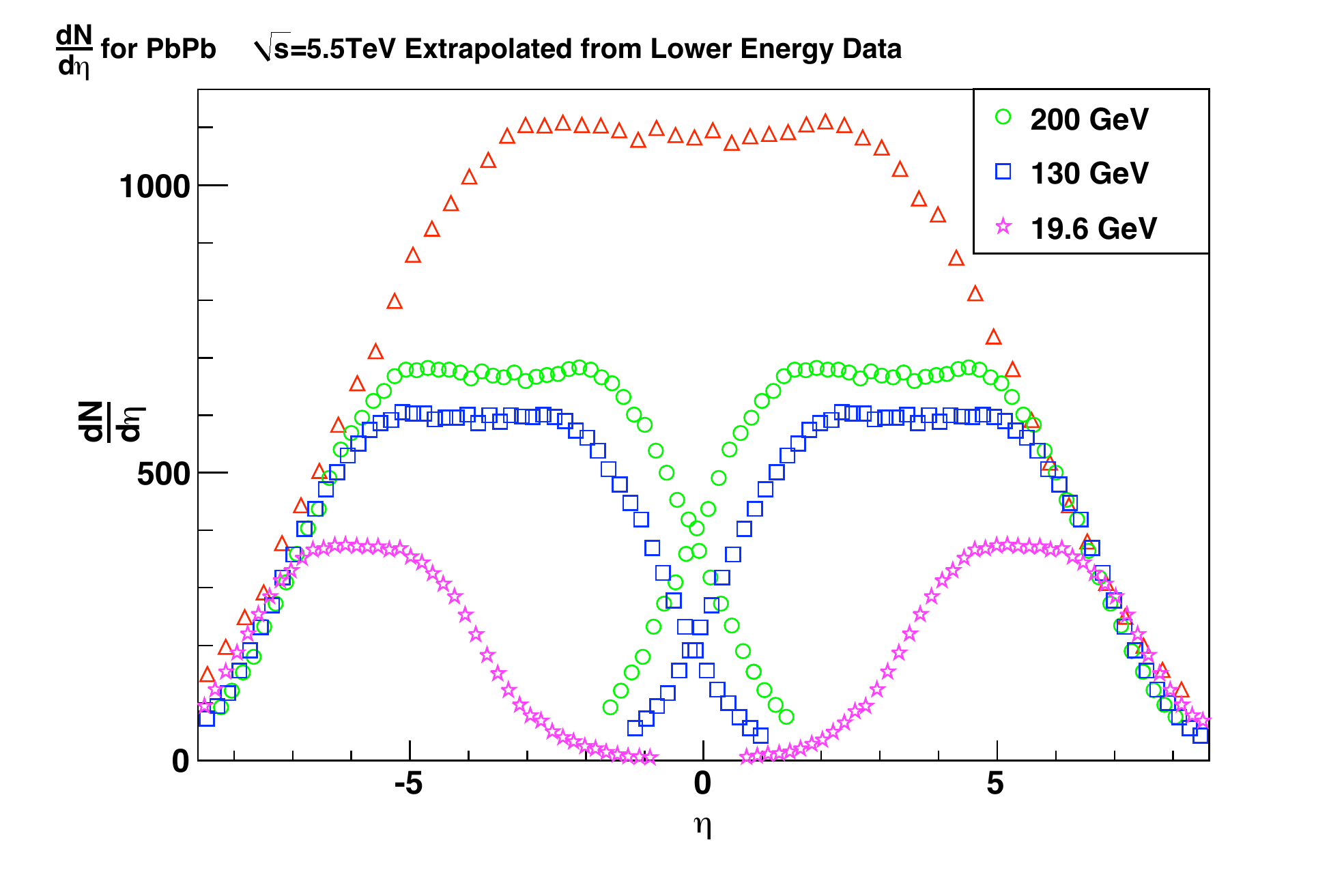}
\label{figure 12}
}
\hfill
\subfigure[]{
\includegraphics[width=.43\linewidth]{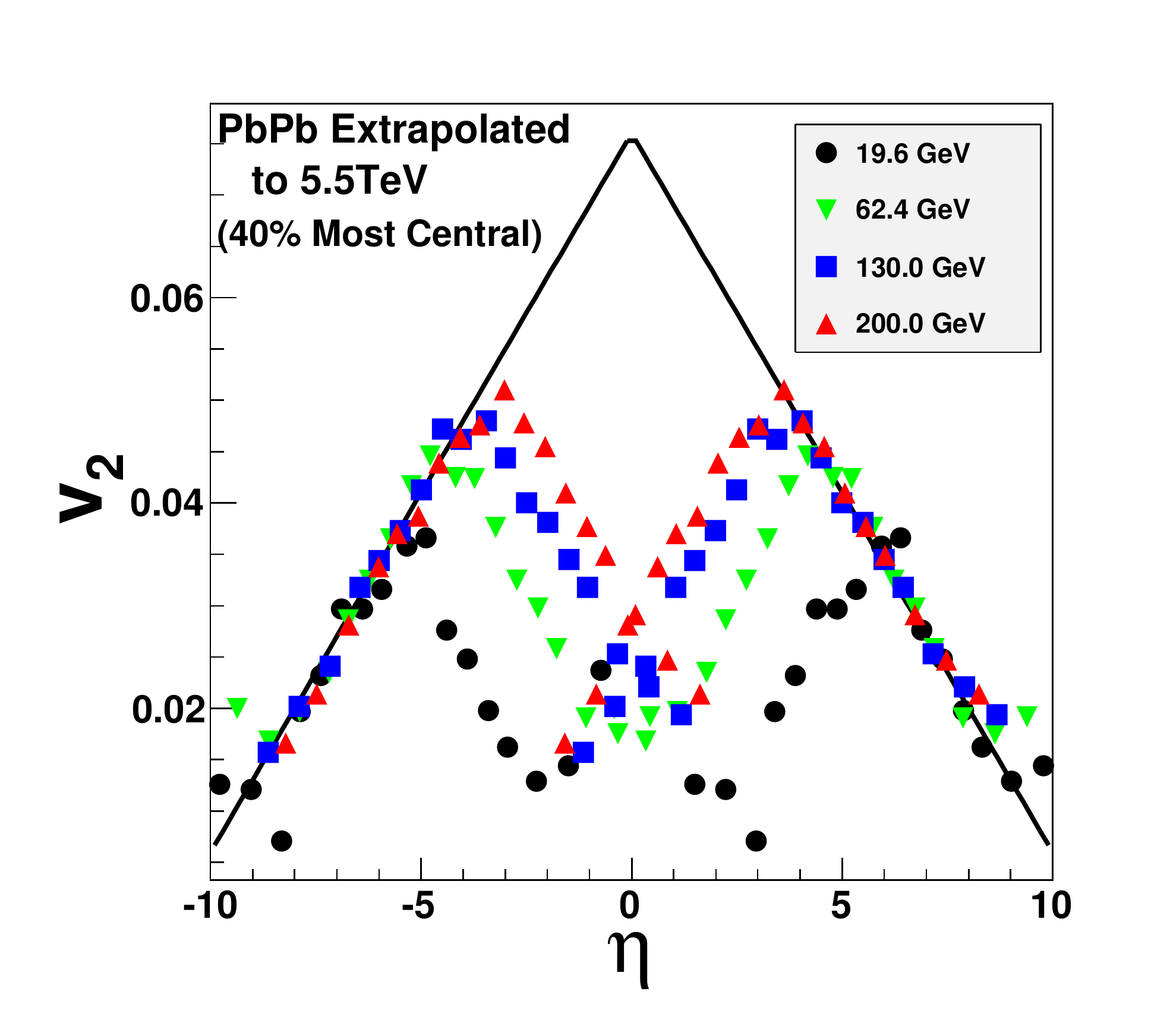}
\label{figure 13}
}
\vspace{-2mm}
\caption[]{\label{figure13}
(Color online) Linear extrapolation of PHOBOS data to LHC energies (5.5 TeV) based on extended longitudinal scaling and logarithmic energy scaling at midrapidity. (a) Pb+Pb pseudorapidity distribution for $N_{part}$=360. (b) Elliptic flow parameter $v_2$ for the 40\% most central Pb+Pb collisions. Figures are from ref. 42.}
\end{figure}


\vspace{-5mm}
\end{document}